# Hydraulic-Assist Driver for Compact Insertion Devices.

Alexander Temnykh[1*], Ivan Temnykh [2] and Eric Banta [1]

[1]CHESS, Cornell University, Ithaca, NY 14850, USA;

[2] Pine Hollow Auto Diagnostics, Pennsylvania Furnace, PA 16865, USA

* Corresponding author, E-mail: abt6@cornell.edu

**Abstract**

We have developed and tested a novel type of driver for insertion devices. In our approach, we compensate for the magnetic forces acting on the insertion device (ID) arrays by installing a number of compact hydraulic cylinders along the ID.

These assisting cylinders are connected to a single hydraulic line and activated by the same pressure. A computer-controlled closed-loop feedback system maintains the target ID array position and an acceptable load on the mechanical driver. This is accomplished by simultaneously controlling both the hydraulic assist pressure and the mechanical driver position. The quantity and location of the compact hydraulic cylinders are optimized to keep the deformation of ID structural components in an acceptable range.

The hydraulic cylinders compensate for over 90% of the strong forces generated by the magnetic field. This reduces the load applied to a conventional mechanical driver to less than 10% of the total load.  The implementation of this type of driver can make insertion devices much more compact without compromising their performance

 In the proof-of-concept experiments, our novel hydraulic-assist driver demonstrated a long-term stability and reproducibility of the ID gap within 2 microns.  Magnetic field stability and reproducibility was within 6e-5 of the normalized field strength or better.

Here we will describe the concept, details of our test setup and the results of the proof-of-concept experiments. We will also discuss the advantages of the hydraulic-assist driver over conventional mechanical drivers, as well as possible applications and future developments.

## 1. Introduction

Conventional insertion devices (IDs) that are used to generate synchrotron radiation in storage rings usually consist of two or more permanent-magnet arrays, which must be positioned with respect to each other with a precision of a few microns. Because the required precision of positioning the arrays is so high, and the magnetic forces between the arrays are very strong (having a nonlinear dependence on the array position), conventional driving mechanisms are complex, massive and expensive.

A number of approaches have been proposed and demonstrated to reduce the magnetic force load on the driving mechanisms.

In [1], the author suggests building four additional magnet arrays to generate a counterforce, and develops a theory to predict the period and strength of these compensating arrays. The prototype demonstrated good consistency with theory, and the load on the mechanical driver was significantly reduced. Further developments of this idea have been described in [2]. Here, to simplify the design, the authors used counterforce magnet arrays formed by multipole monolithic magnets. Experiments showed that this method substantially reduced forces on the mechanical driver and resulted in very satisfactory properties of the ID magnetic field.

A different approach has been proposed in [3]: installing custom-designed conical springs to compensate for the magnetic forces of the ID. The springs exhibited an exponential force-enlargement dependence, similar to the dependence of magnetic forces on the ID gap. This design provided an efficient force compensation for a wide range of ID gaps.

In our approach, we compensate for the magnetic forces acting on the insertion device arrays by installing a number of compact hydraulic cylinders along the ID. These assisting cylinders are connected to a single hydraulic line and activated by the same pressure. We implement a computer-controlled closed-loop feedback system to maintain the target ID array position and an acceptable load on the mechanical driver. This is accomplished by simultaneously controlling both the hydraulic assist pressure and the mechanical driver position. The program monitors the ID gap, load on the mechanical driver, and hydraulic line pressure at all times.

The main motivation for developing this new hydraulic-assist driver is to improve on the design of Cornell Compact Undulators (CCU) [4] by adding a variable-gap option.

Compared to conventional undulators, CCUs are about ten times more compact, much lighter and less expensive. They also require a much shorter fabrication time. Nine CCUs have already been constructed in industry, including two units that have been in operation at the Cornell High Energy Synchrotron Source (CHESS) for four years. Although featuring many advantages, CCUs have two undesirable characteristics. First, they have a ~20% weaker peak magnetic field for the same gap, compared to conventional undulators. Second, the normalized magnetic field errors have a tendency to increase when the magnetic field is adjusted to a low level. These two undesirable effects are partially attributed to the adjustable-phase mode of field control in the CCU, instead of a gap-varying field control, which is the norm in conventional devices.

The compact hydraulic-assist driver presented in this paper has the potential to greatly improve on the existing CCU design by introducing the ability to precisely control the ID gap. The resulting device will inherit the compactness, light weight and fabrication simplicity of the existing CCU, and at the same time, it will have a stronger field and smaller magnetic field errors.

To prove the feasibility of the hydraulic-assist driver concept, we installed the driver prototype on a 30cm-long model of a compact wiggler similar to a CCU and conducted several experiments. In Section 2 we present the design and results of our proof-of-concept experimental study. In Section 3 we summarize the results of the study and discuss future implementation of our prototype hydraulic-assist driver.

## 2. Proof-of-concept experiments

In our proof-of-concept experiments, we used a 2 Tesla Compact Wiggler model constructed a few years ago with the intention to operate it at the Cornell High Energy Synchrotron Source (CHESS) facility. Originally, the model had a fixed ID gap. For the experiments, we modified the Wiggler model by installing our prototype hydraulic-assist driver to vary the ID gap.

In our experiments we addressed two potential concerns of using a hydraulic-assist system for controlling the ID array position: First, long-term magnetic field stability in steady state; second, magnetic field reproducibility after adjusting the ID gap.

Both of these potential concerns are due to the unknown behavior and performance of the hydraulic components in the system. System leaks, hysteresis in the pressure transducer signal, stiction in the hydraulic cylinders, etc., may cause unpredictable fluctuations of hydraulic pressure and compensating forces and thus compromise the stability of the ID gap thus the magnetic field.

### 2.1 Model description

The magnetic structure of the experimental model (Fig.1) consisted of NbFeB permanent magnet blocks and vanadium poles. At a minimum ID gap of 7mm, peak field was ~1.9T and magnetic forces between magnet arrays were ~ 492 kg-force (1084 lbs-force).

Originally, both upper and lower magnet arrays had been clamped down to the frame, but for the varying-gap demonstration, the upper array was released and made movable in the vertical direction. The gap-varying mechanism included both hydraulic (main) and mechanical (trimming) drivers.

The hydraulic driver consisted of six miniature hydraulic cylinders (Vektek, model 20-01-07). Each cylinder had a piston with 0.11 sq. inch area, 0.75" stroke, 550 lb. load capacity and a 5/8" DIA threaded body. All cylinders were connected to a single hydraulic line equipped with a pressure transducer (WIKA, Type A-10 / 0 - 3000 psi). Cylinder bodies were attached to the frame (3 on each side) using aluminum bars as shown on Fig.2a. The cylinder rods pushed vertically on the upper array extensions. The number of cylinders was chosen to keep the required hydraulic pressure below 2000 psi. For that pressure range, a large variety of hardware and fittings is available in local automotive or hardware stores. In our case, 2000 psi of hydraulic pressure fed to 6 cylinders provided a maximum force of 2000*0.11*6 = 1320 lbs- force.  This exceeded the required 1084 lbs-force by a reasonable 20% margin.

As a hydraulic pressure source, we used a hydraulic ram pump (Pittsburgh Hydraulics 10,000-PSI model) driven by a linear actuator (Kollmorgen, model EC2), shown in Fig. 2b.

The mechanical driver consisted of two custom compact z-stages with a few hundred lbs. capacity each (5.1 and 5.2 in Fig.3) and a "flex" plate (7) attached to the frame by four rods (8). The two z-stages were placed between the "flex" plate and the upper magnet array (one on each end of the model) and were clamped down to both (see Fig.3). By moving the two z-stages we applied additional forces between the upper array and the frame.  Two strain gauges from Omega Engineering, glued to the "flex" plate in proximity of the z- stages, monitored local strain proportional to the z-stage load.

Two digital indicators (Fowler-Sylvac Mark VI) (3.1 and 3.2) with ~1-micron resolution monitored the upper array displacement with respect to the frame (ID gap variation) on both ends of the model.

To drive the z-stages and the linear actuator governing the hydraulic pump, we used 3 channels of a stepping motor controller (Galil Motion Control DMC_2182).

### 2.2 Operation principle

In steady state, there is a balance between magnetic, hydraulic and mechanical forces applied to the upper array. It can be expressed as:

$$F_{mag}(g) + F_{hydr}(P) + F_{stage} = 0$$

Here: $F_{mag}$ - magnetic forces between magnet arrays (dependent on ID gap $g$); $F_{hydr}$ - force exerted by the hydraulic cylinders; $F_{stage}$ -force applied by the z-stages. In this case, $F_{hydr} = A \times P$, where $A$ is a total piston area and $P$ is the hydraulic pressure which controlled by the pump and monitored by the pressure transducer.  $F_{stage}$ was monitored through the two strain gauges glued to the "flex" plate..

In the beginning, we measured the dependence of the required hydraulic pressure ($P$) on ID gap ($g$) while keeping the strain of the "flex" plate, i.e.  $F_{stage}$, close to zero. . For each step, we changed the ID gap by moving the z-stages. After that, the hydraulic pressure was adjusted by the pump to have zero strain on the "flex" plate ($F_{stage} = 0$).  In this way, at each point we achieved a balance between the magnetic and hydraulic forces. The measured and expected dependencies of hydraulic pressure on ID gap are presented in Fig.4.  Solid line shows the measurement; dashed line presents data predicted by 3D magnetic modeling of the system using OPERA software. The measurement and prediction are in good agreement.

To control the ID gap, we developed a program using LabView software. The program executed two independent feedback loops at a 1Hz rate.

In the first loop, the program was monitoring the upper array position by reading the digital indicators while moving the z-stages in small steps toward the target. Before each step, the program was checking the strain gauges, i.e. z-stages load. The motion was activated only if the load was "low". The "low load" condition was satisfied only when the lift force created by the hydraulic cylinders entirely compensated for the attractive magnetic forces between the arrays.

The second loop controlled the hydraulic pressure. The program read the digital position indicators, calculated the ID gap and, using data presented on Fig.4, calculated the hydraulic pressure needed for ~95% compensation of the magnetic forces. The desired pressure was compared with the actual (measured by the pressure transducer) and, if the desired pressure was higher, the program activated the hydraulic pump to add oil into the system. If the desired pressure was lower than the actual, the program just cycled idly until the actual pressure was reduced due to leaks.

### 2.3 Long-term stability check

The reason for the long-term stability concern was the assumption that small leaks in a hydraulic system can create difficulties in maintaining a stable pressure for a long time, which is required for a steady ID gap. To address this problem, we recorded several critical parameters such as magnetic field, hydraulic pressure, upper array position etc., for 120 hours (5 days) at a 1 Hz rate while maintaining a 13 mm ID gap. Fig. 5 depicts the results.

The plot in Fig. 5a shows the 120 hours (5-day) record of both ends of the upper array position. The target position was 5 mm. The data indicates no more than 2 microns deviation from the target. Taking into account a 1-micron gauge resolution and the +-2-micron dead band used in the feedback loop, a +-2-micron deviation from the target is quite satisfactory.

Fig. 5b presents the hydraulic system pressure recorded over the 120-hour time frame. The data indicates a 654.58 psi average pressure with a standard deviation of 4.38 psi. The deviation is consistent with the given transducer performance specifications: ~3 psi repeatability and 3 psi drift.

Note that both parameters, end positions and pressure, are in fact the signals kept constant by feedback loops. Thus, if these signals are drifting due to sensors imperfection, the measured parameters will not agree with actual, but the recorded data will not show that.

The most indicative parameter is the actual magnetic field in the ID gap, plotted in Fig. 5c along with permanent magnet material temperature. The temperature data (left scale) shows 22.99 °C average temperature with 2 °C peak-to-peak variation, correlated with the time of day. The cause of the variation was ambient temperature changes in the building. Magnetic field data shows a 14021.04 Gauss average field and approximately 45 Gauss peak-to-peak variation. The latter is strongly correlated with temperature.

To evaluate this correlation, we plotted magnetic field as a function of temperature (Fig.5d). The plot reveals a linear dependence and the fit indicates -21.75 Gauss or -0.15% field variation per one degree C of temperature change. This well agrees with what one can expect taking into account a -0.12% contribution from NbFeB PM magnetization reduction and -0.024% from ID gap increase due to aluminum frame thermo-expansion, giving a -0.144% total field variation. Now, knowing how the field depends on temperature, we can correct the raw field data to project it to the condition of a constant (23 °C) ambient temperature. The result of the projection, together with the original record, are plotted on Fig. 6. The projected data has much less variation: It shows a 14021.14- Gauss average field with just 0.91 Gauss standard deviation. That translates to (at least) 0.65e-4 of normalized magnetic field stability.

Such stability will satisfy the most demanding applications. This small deviation of the field also validates the stability seen in the record of the upper array position and the hydraulic pressure.

### *2.4 Repeatability check*

In the repeatability test, we cycled the upper array position between "zero" and 5 mm in respect to the frame. That corresponded to the ID gap changing between 7 and 13 mm. At each position we dwelled for ~10 minutes while recording position, field, temperature, etc. Fig. 7 illustrates the process.

Fig. 7a shows a periodic change of the upper array position and, as expected, it is correlated with the magnetic field strength. Fig. 7b shows three data sets: raw temperature signal from a thermistor located in the wiggler gap, the temperature averaged over the dwell time, and the magnetic field. The thermistor raw data is quite noisy, but it reveals a slow temperature drift of about 0.25 °C during the experiment. The temperature data averaged over the dwell time confirms this drift and, in addition, reveals a dependence of the thermistor signal on ambient magnetic field. Comparing averaged temperatures for "zero" and 5mm upper array positions, corresponding to ~2kG and 1.4kG field in the ID gap, one can see a fast ~0.1 °C temperature signal change. This change cannot be physical, because of the significant model mass and absence of fast-varying heat sources. Most likely, it is caused by the dependence of thermistor characteristics on the magnetic field.

In the following analysis, we used the temperatures averaged over the dwell time to correct for the effect of permanent magnet material magnetization change due to temperature drift.

Figure 8 shows the field for all three cycles in fine scale: a) at "zero" position and b) at "5mm". Boxes on top display average temperatures and average field during the dwell time.

On both plots in Fig. 8, one can see a systematic drift of the field and average temperature from one cycle to another. Knowing how the field depends on temperature (see section 2.3), we calculated the field that we would measure at a constant 23 °C temperature. The data, together with the results of statistical analysis, are summarized in Table 1. The raw data analysis indicated 1.33 and 1.43 Gauss RMS field repeatability, corresponding to 6.6e-5 and 1.0e-4 relative field change. The analysis of the data which was corrected for temperature variation, showed 1.07 and 0.54 Gauss RMS of the field variation corresponding to *5.3e-5* and *3.9e-5* relative field change. This high level of repeatability will satisfy the most demanding applications.

## 3   Discussion and Conclusion

We have proposed a new concept and experimentally demonstrated the feasibility of an insertion device driver based on hydraulic assistance.

In this scheme, the hydraulic system compensates for most (~90-95%) of the magnetic forces, while the mechanical driver compensates for the remaining force (10-5%). In the proof-of-concept experiments, prototype of hydraulic-assist driver showed excellent performance. A long-term field stability of 0.6e-4 and field repeatability of 0.5e-4 was demonstrated.

In comparison with methods described in [1, 2, 3], the hydraulic assistance method for magnetic forces compensation does not require any special or custom-designed components. All parts are very inexpensive and readily available off-the-shelf. In addition, our method is much more flexible. By varying program parameters, it is easy to change the level of load on the mechanical drive, adjust the hydraulic compensating counterforces as a function of the ID array position, optimize dynamic properties, etc.

As a next step in the development and implementation of this concept, we are considering two projects:

The first project is to modify an existing 1.5m-long CCU (see [4]) by adding hydraulic assistance to the existing mechanism for array phase adjustment. Obtained experience will help us design and build longer and more efficient versions of existing types of adjustable-phase IDs such as CCUs and Delta undulators [5,6].

In the second project, we plan to convert another CCU into a varying-gap undulator by adding a hydraulic-assist driver similar to the one we have used in the described experiments.

In conclusion, we believe that further development and application of this new hydraulic-assist driver will help to design and build a new generation of long, lightweight, cost-efficient compact insertion devices with both, variable-gap and adjustable-phase mechanisms for field control.

## 4   Acknowledgment

Authors would like to thank Dr. Svetlana Temnykh for useful discussions and help with writing this paper and Nahla Minges for proofreading.

The work was supported by the NSF award DMR-1332208.

## *Figures*

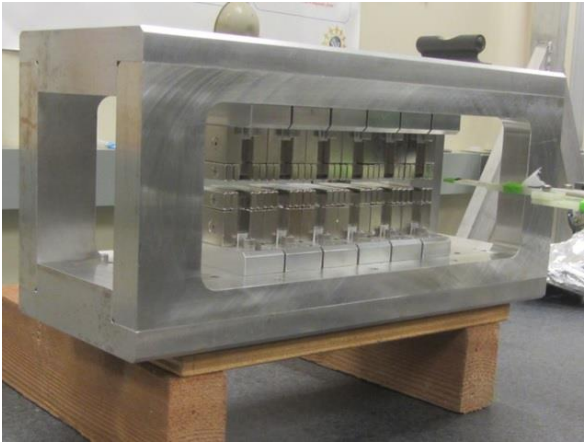

| Magnetic structure Length [mm] | 291.9 |
|---|---|
| Min Gap [mm] | 7 |
| Peak Field [T] | 1.9 T |
| Number of poles | 6 |
| Period [mm] | 73 |
| Magnetic Forces between arrays at minimum gap [kg-force / lbs.-force] | 492 / 1084 |

*Figure 1, Compact Wiggler model used in experiments.*

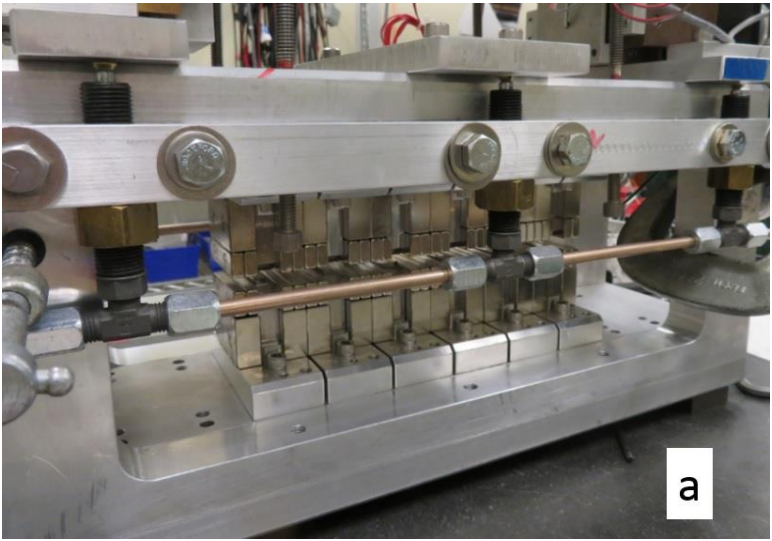
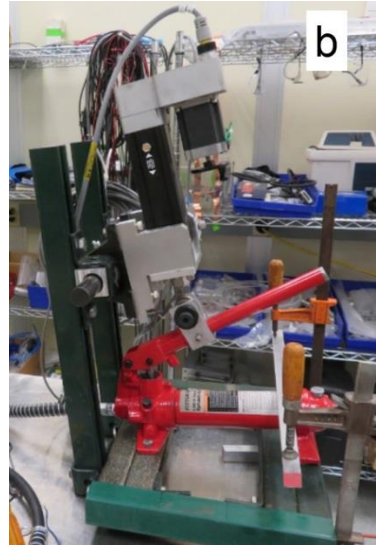

*Figure 2. Hydraulic driver components. a) - Hydraulic cylinders attached to the frame. b) - hydraulic pressure source consisted of "Pittsburgh 10000 PSI Hydraulic ram pump" and linear actuator.*

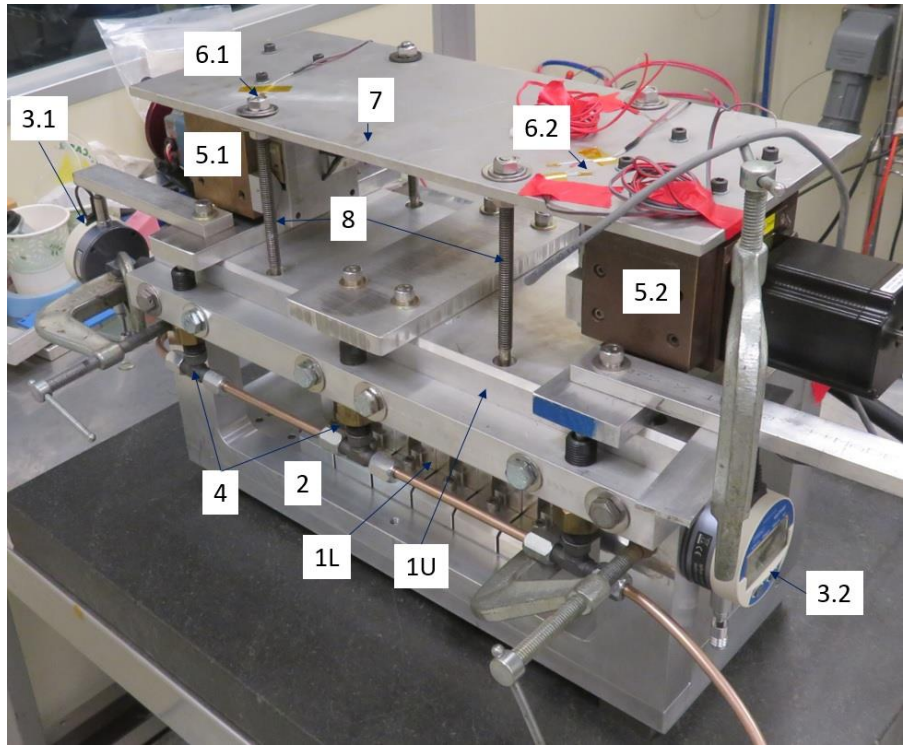

*Figure 3. Compact wiggler model fitted with hydraulic assistant varying gap driver. Here: 1L and 1U – lower and upper magnet arrays; 2 - frame; 3.1 and 3.2 digital indicators measuring upper array displacement in respect to frame; 4 – hydraulic cylinders; 5.1 and 5.2 – z-stages; 6.1 and 6.2 – strain gauges; 7 – "flex" plate; 8 – rods coupled "flex" plate with frame.*

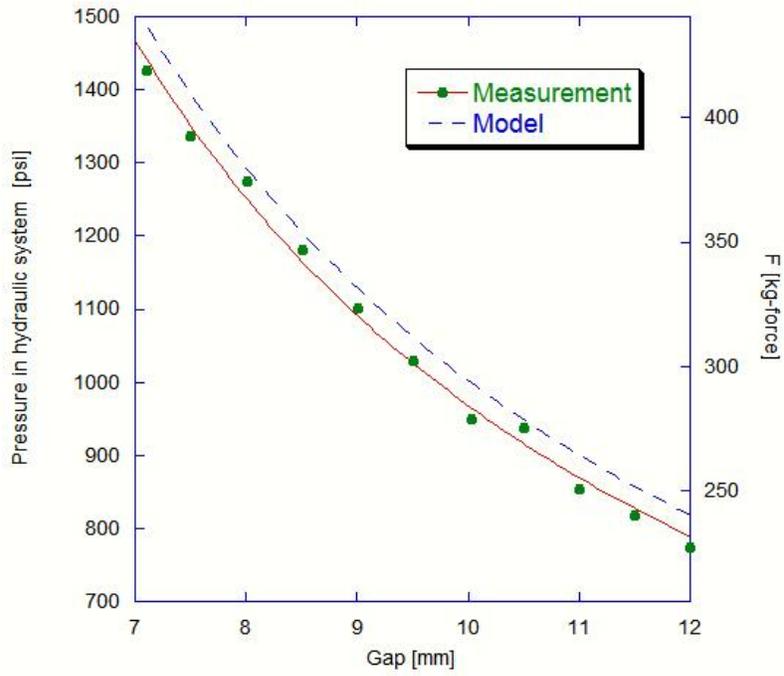

*Figure 4. Hydraulic pressure (left scale) and total hydraulic actuators forces (right scale) balancing magnetic forces as function of gap.*

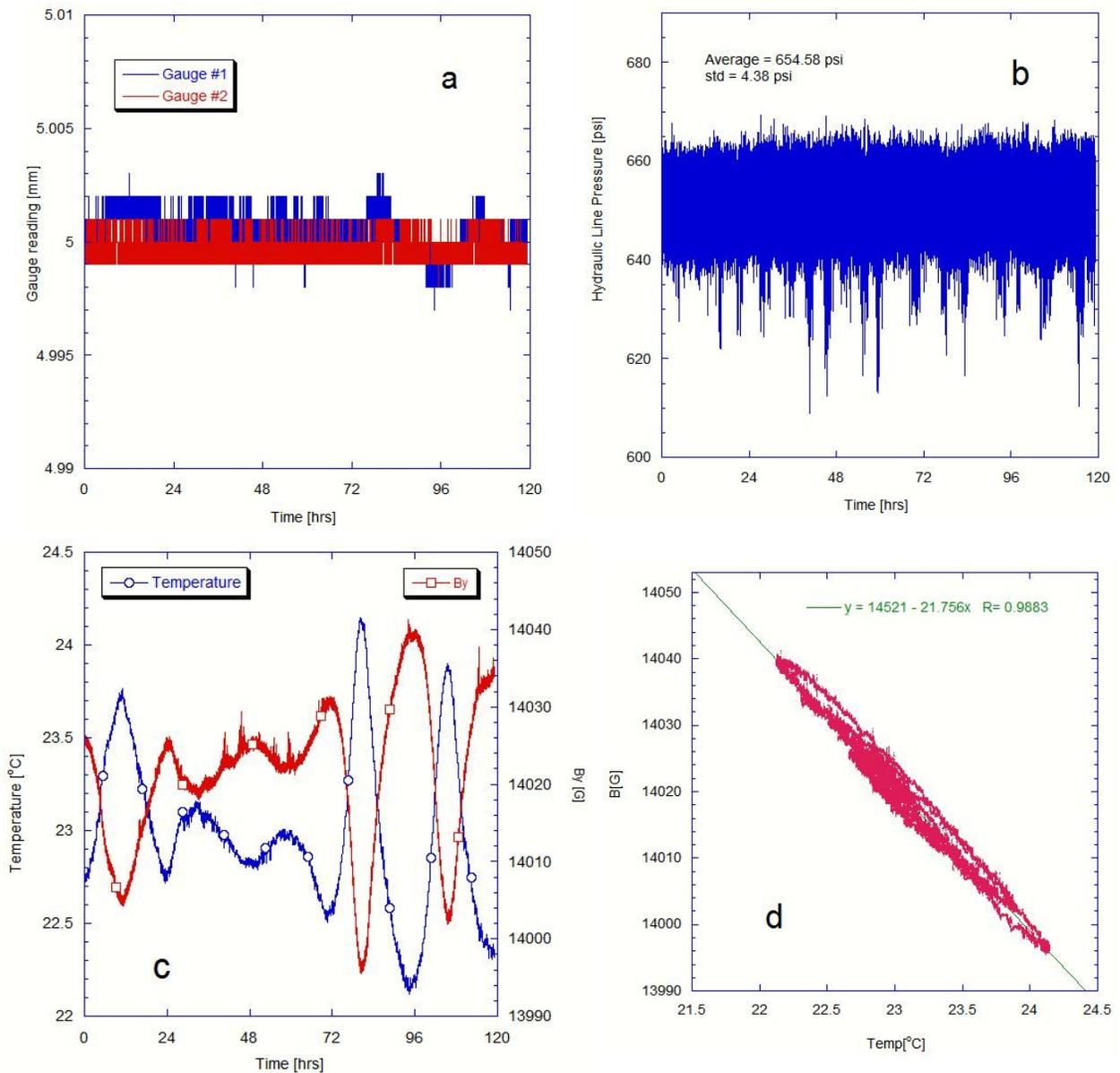

*Figure 5, 120 hour steady gap record. a) - upper array position measured with gauges; b) - pressure in hydraulic system; c) - magnetic field and temperature in the gap measured with Hall Probe and thermistor; d) – magnetic field plotted versus temperature*

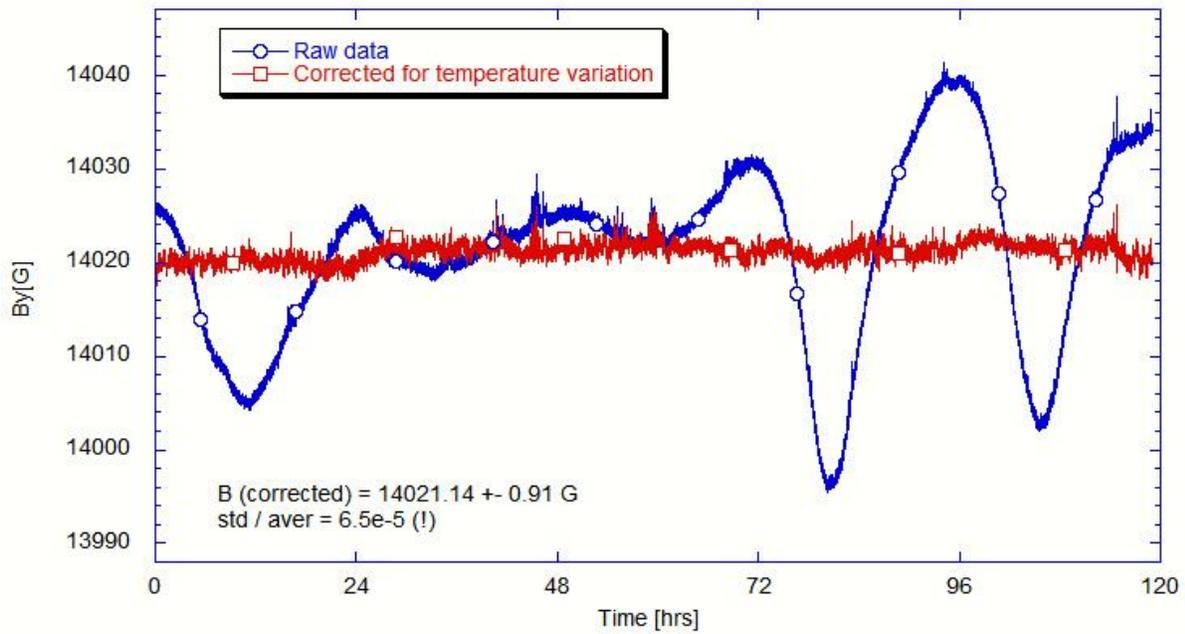

*Figure 6. Raw and temperature corrected magnetic field record.*

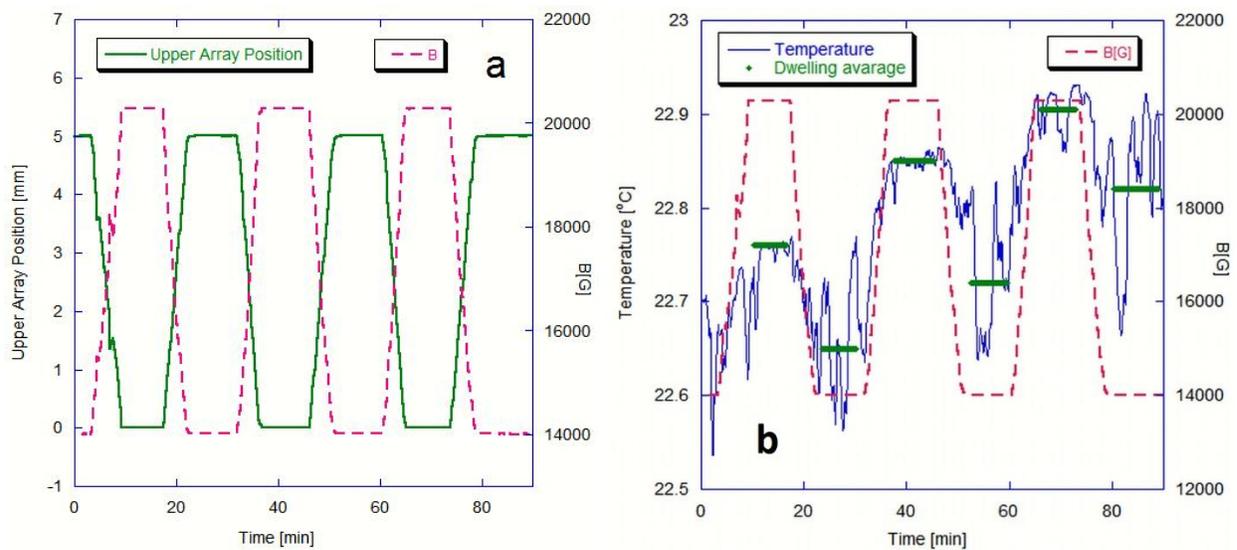

*Figure 7. Records illustrating repeatability study experiments. On the left - upper array position (solid line) and the gap magnetic field (dashed line), on the right - temperature record and temperature averaged over dwelling (solid lines) and the gap magnetic field record (dashed line).*

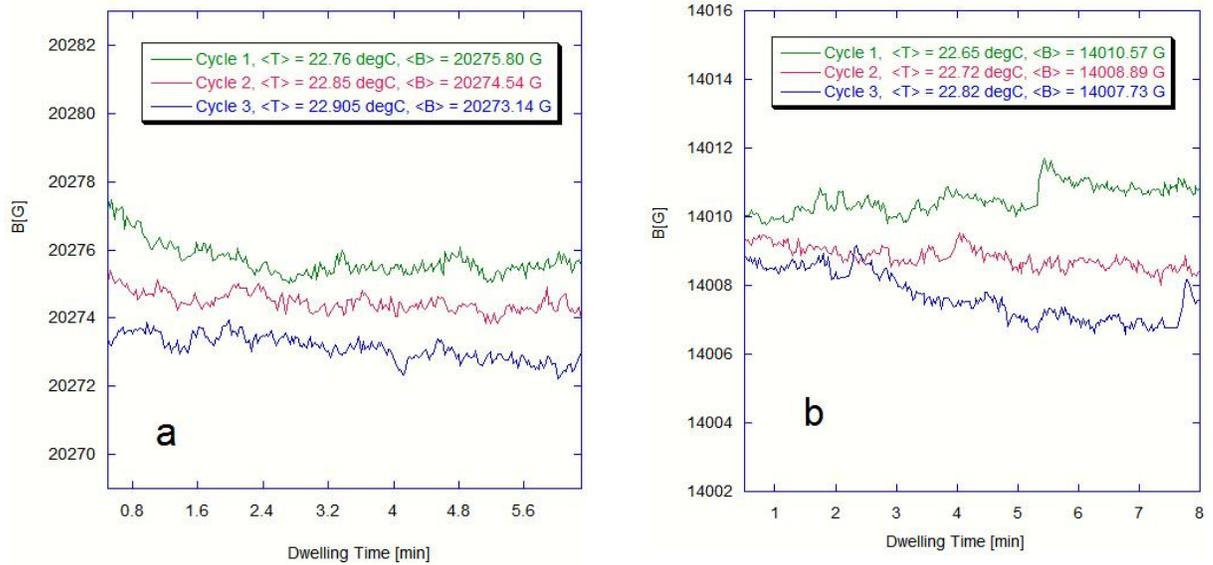

*Figure 8. Field record for three cycles: a) - at "zero" and b) - at "5mm" (on the right) upper array position.*

*Table 1 Measured field and field projected to 23°C.*

|          | "Zero" point |         |              | "5mm" Point |          |              |
|----------|------|----------|--------------|------|----------|--------------|
| Cycle #  | <T>  | <B> raw  | <B> @ 23degC | <T>  | <B> raw  | <B> @ 23degC |
| 1        | 22.76 | 20275.8  | 20268.26     | 22.65 | 14010.57 | 14002.97     |
| 2        | 22.85 | 20274.54 | 20269.83     | 22.72 | 14008.89 | 14002.81     |
| 3        | 22.91 | 20273.14 | 20270.31     | 22.82 | 14007.73 | 14003.82     |
| <B> =    |      | 20274.49 | 20269.47     |      | 14009.06 | 14003.20     |
| std =    |      | 1.33     | 1.07         |      | 1.43     | 0.54         |
| std/<B> = |     | 6.6E-05  | **5.3E-05**  |      | 1.0E-04  | **3.9E-05**  |